# Quantum Wavemetry via the *Mth*-Power Unitary of a Mach-Zehnder Interferometer

Byoung S. Ham[1,2,3]

[1]Department of Electrical Engineering and Computer Science, Gwangju Institute of Science and Technology, 123 Chumdangwagi-ro, Buk-gu, Gwangju 61005, South Korea
[2]Qu-Lidar, 123 Chumdangwagi-ro, Buk-gu, Gwangju 61005, South Korea
[3]Department of Electrical and Computer Engineering, Oregon State University, Corvallis, OR 97331, USA
(March 05, 2026; bham@gist.ac.kr)

**Abstract**
A quantum wavemetry scheme based on the coherence de Broglie wavelength (CBW) is proposed using an $M$-coupled Mach–Zehnder interferometer (MZI) architecture to achieve superresolution sensing and metrology. Although CBW does not attain the Heisenberg limit, it circumvents the key practical limitations of N00N-state-based quantum sensing, including restricted photon number $N$, reduced fringe visibility, and strong susceptibility to photon loss. The CBW approach enables loss-tolerant operation with arbitrarily large $M$, while maintaining near-unity fringe visibility. Fully compatible with coherence optics, the CBW scheme can be directly integrated into conventional wavemetry systems, providing both superresolution and enhanced sensitivity. A proof-of-principle experiment demonstrating CBW-based superresolution is implemented using a Sagnac-integrated round-trip MZI structure for $M = 2$, validating the feasibility of the proposed quantum wavemetry design.

Quantum sensing and metrology have been intensively studied over the last few decades to overcome the fundamental limitations of classical counterparts [1-7]. According to Fisher information theory, the minimum uncertainty in estimating an unknown parameter is bound by the Cramer-Rao lower bound (CRLB) [8]. In classical measurement systems employing independent resources, the CRLB yields the shot-noise limit (SNL), also known as the standard quantum limit, which originates from the Poissonian statistics of uncorrelated particles [7]. The SNL corresponds to the $1/\sqrt{N}$ scaling of the CRLB for parameter estimation using $N$ independent (classical) resources [8]. By employing nonclassical states of light, quantum sensing can surpass the SNL and approach the Heisenberg limit in phase estimation [1-7]. Thus, quantum sensing and metrology can improve phase sensitivity from the classical $1/\sqrt{N}$ scaling to the Heisenberg scaling of $1/N$ [6]. N00N-state-based quantum sensing exhibits not only supersensitivity ($1/N$) but also superresolution [7]. This superresolution arises from the photonic de Broglie wavelength (PBW) [9], given by $\lambda_B = \lambda_0/N$, which is fundamentally different from classical high-resolution effects based on multi-wave interference [10]. When generated via spontaneous parametric down-conversion (SPDC), N00N states arise from a probabilistic process with a super-Poissonian photon number distribution [11]. Consequently, the generation efficiency of N00N states decreases exponentially with increasing photon number $N$ [4,11], and practical implementations typically rely on post-selection of specific $N$-photon events [12]. Experimentally, the maximum demonstrated photon number for PBW interference remains limited (e.g., $N = 18$) [13], far below the regime where scalable quantum advantage becomes practically significant [14]. For these reasons, the practical deployment of N00N-state-based quantum sensing remains challenging.

Classical sensing and metrology [15-18] rely on coherence and interference implemented in interferometric platforms such as Mach-Zehnder interferometers (MZIs) [19], Michelson interferometers [20], and Fizeau-type interferometers [21]. To overcome the diffraction limit, both linear [15-21] and nonlinear optical approaches [22,23] have been explored. In linear optics, resolution enhancement beyond the diffraction limit can be achieved through multi-wave interference [10]. According to the definition of Fisher information [8], sensitivity can also be improved by increasing the acquisition time of the measurement outcomes [24]. For this purpose, coherent rephasing among inhomogeneously broadened atoms (spins) has been applied in techniques such as the Ramsey fringes [23] and spin echoes [24] in the nonlinear optics regime. A Fizeau-type interferometer utilizes both transverse and longitudinal interference to extract unknown parameters rapidly and efficiently [21]. Although the resolution of classical sensor platforms can be coherently enhanced through coherent effects to achieve high resolution beyond the diffraction limit, the sensitivity scaling remains limited by the SNL. The corresponding Fisher information must properly account for all measurement resources [8]. In that sense, classical linear and nonlinear optical sensing platforms cannot surpass the SNL, where the apparent $1/N$ scaling arising from cumulative phase contributions does not imply sub-SNL performance unless it exceeds the scaling permitted by the total resource budget.

Recently, alternative sensing strategies have been proposed to overcome practical limitations in both classical and quantum sensing [25-28]. One approach employs intensity-correlation measurements combined with post-selection techniques [27,28], which are widely used in quantum sensing [1-7]. Another approach utilizes the *M*th power of the MZI unitary operator implemented in a cascade MZI architecture [25,26,29]. The former relies on the quantum eraser concept [30], one of the fundamental mysteries of quantum mechanics [31]. The latter is referred to as the coherence de Broglie wavelength (CBW), which emerges from coherently coupled MZIs [25]. Although implemented using classical coherence optics, the physical origin of CBW differs fundamentally from conventional high-resolution techniques, including multi-wave interference in linear optics [15-21] and population manipulation in nonlinear optics [22-24]. Thus, CBW is conceptually distinct from classical high-resolution methods and instead relates to PBW-like superresolution, analogous to that observed in quantum sensing. In addition to PBW-like superresolution, CBW also enables enhanced phase sensitivity [29]. Here, CBW is incorporated into a conventional wavemeter, specifically a Fizeau-type interferometer [21], to realize a CBW-based quantum wavemeter. Owing to its compatibility with coherence optics, the proposed scheme avoids the stringent constraints associated with N00N-state-based quantum sensing. In terms of scalability, superresolution with $M = 100$ is, in principle, achievable using CBW (discussed elsewhere). Furthermore, the phase sensitivity of the proposed quantum wavemeter is enhanced by a factor of *M* relative to the standard quantum limit under equivalent measurement conditions.

Figure 1 shows a schematic of the proposed quantum wavemeter. A conventional sensing system based on a single MZI is modified to realize the CBW using an anti-symmetrically coupled MZI architecture through a dummy MZI ($\psi$). For the special case of $M = 2$, a Sagnac interferometer-based CBW configuration is illustrated in the lower inset, where the two coupled MZIs are folded into a round-trip propagation scheme. For a scalable *M*th order quantum wavemeter, the inset configuration can be extended to larger *M* (discussed elsewhere). As analyzed quantum mechanically [29], the CBW architecture in Fig. 1 provides an *M*-fold enhancement in both resolution and phase sensitivity compared with a conventional single-MZI configuration [21,32].

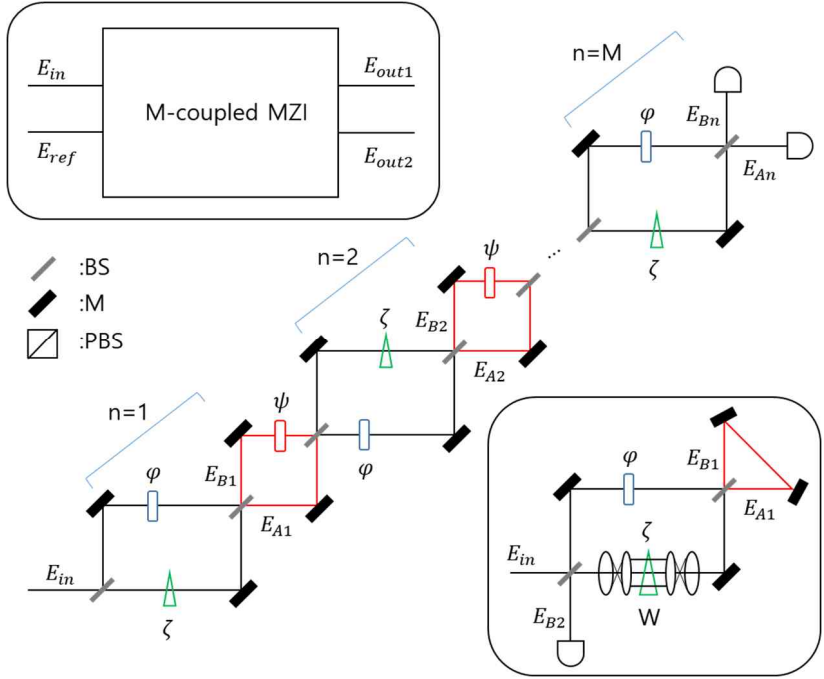

**Fig. 1.** Schematic of a quantum wavemeter. Inset: a block diagram. $\varphi$ and $\zeta$ are longitudinal and transverse phases, respectively. $\psi$ is the coupling phase. BS: beam splitter, M: mirror, PBS: polarizing BS, W: wedge optics.

In Fig. 1, the anti-symmetrically coupled MZI structure is indexed by n=1, 2,..., M, where each unit MZI block contains a phase parameter $\varphi$, representing the unknown quantity to be estimated. The parameter $\zeta$ controls the transverse coordinate used to implement the Fizeau interferometer configuration in a conventional wavemeter. The consecutive MZIs are arranged in a perfectly anti-symmetric configuration by alternately swapping the MZI paths, while $\varphi$ governs the axial (longitudinal) phase information. This anti-symmetrically coupled MZI architecture enables the effective $M$th power of the MZI unitary operator [29]. The apparent anti-symmetry between adjacent $\varphi$-MZIs ensures that the same path basis of the MZIs is preserved through the intermediate $\psi$-MZI, thereby realizing the $M$th power of the MZI unitary transformation [29].

Neglecting $\zeta$, the output intensities of the $M$-coupled MZIs in Fig. 1 can be derived coherently as [25,26]:

$$I_{An} = I_0 \cos^2\left(\frac{M\varphi}{2}\right), \quad (1)$$

$$I_{Bn} = I_0 \sin^2\left(\frac{M\varphi}{2}\right), \quad (2)$$

where $I_k = E_k E_k^*$. According to the Rayleigh criterion, Eqs. (1) and (2) exhibit superresolution with fringe spacing $\Delta\varphi = \pi/M$, analogous to the PBW in quantum sensing [3-7]. Unlike the efficiency reduction proportional to the power of $N$ in N00N-state-based quantum sensing [11-13], the CBW configuration maintains near-unity efficiency independent of $M$, owing to the intrinsic $M$th-order filtering process in the coupled interferometer architecture [29]. With respect to the SNL, the random variable in the CBW measurement corresponds to the detected output intensity characterized by the mean photon number $\mu$. Consequently, the phase uncertainty scales as $\Delta\varphi_{CBW} \propto \frac{1}{M\sqrt{\mu}}$, indicating an $M$-fold improvement in sensitivity relative to a conventional single-MZI measurement, while remaining consistent with SNL scaling [29]. This deterministic enhancement is analogous to sensitivity improvements obtained through extended acquisition time or coherent rephrasing techniques [23,24].

From a fully quantum mechanical description, the path projectors of a single MZI in Fig. 1 can be written as $\Pi_u = |u\rangle\langle u|$ and $\Pi_l = |l\rangle\langle l|$, where $u$ and $l$ denote the upper and lower interferometer paths, respectively. The path-population difference operator is therefore $\sigma_z = \Pi_u - \Pi_l$, which corresponds to the which-path observable of the interferometer. In this representation, the MZI operates on a path qubit. With the phase shift $\varphi'$ introduced in Fig. 1, where $\varphi' = \varphi - \zeta$, the phase operator is given by $P_u(\varphi') = e^{i\varphi'}|u\rangle\langle u| + |l\rangle\langle l|$. Introducing a global phase $\varphi'/2$, the unitary transformation of the MZI can be expressed using Pauli operators as:

$$U_{MZI}(\varphi) = e^{i\varphi'/2}\begin{bmatrix} \cos\left(\frac{\varphi'}{2}\right) & i\sin\left(\frac{\varphi'}{2}\right) \\ i\sin\left(\frac{\varphi'}{2}\right) & \cos\left(\frac{\varphi'}{2}\right) \end{bmatrix}, \quad (3)$$

where the no-loss beam splitter and phase operators are represented by $\frac{1}{\sqrt{2}}\begin{bmatrix} 1 & i \\ i & 1 \end{bmatrix}$ and $\begin{bmatrix} e^{i\varphi} & 0 \\ 0 & e^{\zeta} \end{bmatrix}$, respectively.

For the anti-symmetrically coupled MZI chain, the effective unitary transformation corresponds to the $M$th power of the single-MZI operator:

$$U_{CBW}(\varphi') = (U_{MZI}(\varphi'))^{\otimes M} = e^{iM\varphi'/2}\begin{bmatrix} \cos\left(\frac{M\varphi'}{2}\right) & i\sin\left(\frac{M\varphi'}{2}\right) \\ i\sin\left(\frac{M\varphi'}{2}\right) & \cos\left(\frac{M\varphi'}{2}\right) \end{bmatrix}. \quad (4)$$

Equation (4) describes the CBW superresolution, which has the same functional form as PBW interference in N00N-state-based quantum sensing [3-7]. Unlike PBW, which relies on $N$-photon entangled states in a single MZI [3-7], CBW arises from coherent phase correlations among $M$ cascaded MZIs. The dummy $\psi$-MZI (red) in Fig. 1 plays a crucial role in preserving the path basis of MZIs, satisfying $D^\dagger \sigma_z D = \sigma_z$, where $D$ denotes the dummy MZI, satisfying $[D, \sigma_z] = 0$. Consequently, the CBW unitary transformation can also be written as $U_{CBW}(\varphi') = \left(e^{i\varphi'\sigma_y/2}\right)^M = e^{iM\varphi'\sigma_y/2}$, which directly yields the $M$-fold phase multiplication responsible for CBW superresolution.

For the Fisher information analysis, the interferogram of the $\zeta$-caused fringe is modeled as:

$$y_k = \mu\left(a + b\cos\left(2\pi f_M x_k + \phi\right)\right) + n_k, \quad (5)$$

where $n_k \sim \mathcal{N}(0, \sigma^2)$ denotes independent Gaussian intensity noise, $f_M$ is the spatial fringe frequency, $b$ is the visibility, and $\mu$ is the mean intensity. In the CBW configuration, the effective frequency scales as $f_M = Mf$, where $M$ is the number of anti-symmetrically coupled MZIs. Assuming $a$, $b$, and $\phi$ are known, the log-likelihood function is Gaussian, and the Fisher information for estimating $f$ is:

$$I(f) = \frac{1}{\sigma^2}\sum_{k=0}^{m-1}\left(\frac{\partial y_k}{\partial f}\right)^2. \quad (6)$$

Since $f_M = Mf$, the derivative becomes $\frac{\partial y_k}{\partial f} = -b\mu(2\pi M x_k)\sin(2\pi f_M x_k + \phi)$, yielding $I(f) = $

$\frac{b^2\mu^2 M^2 (2\pi)^2}{2\sigma^2}\sum_k x_k^2$.

For $m$ spatial sampling points $x_k$ distributed approximately uniformly along the wedge length $L$, the second moment can be approximated as $\sum_k x_k^2 \sim mL^2/3$. Substituting this into the Fisher information yields $I(f) \sim \frac{b^2\mu^2 M^2 mL^2 (2\pi)^2}{6\sigma^2}$. The corresponding CRLB for estimating $f$ becomes:

$$\text{var}(\hat{f}) \geq \frac{6\sigma^2}{b^2 M^2 \mu^2 (2\pi)^2 mL^2}. \tag{7}$$

Defining the fringe count $K_M = f_M L = M f L$, the corresponding uncertainty is $\text{std}(\hat{K}_M) \sim \frac{\sqrt{6}\sigma}{2\pi b M \mu \sqrt{m}}$. Since the wavelength estimate satisfies $\Lambda_M = L/K_M$, the fractional uncertainty becomes:

$$\text{std}\left(\frac{\hat{\Lambda}_M}{\Lambda_M}\right) \geq \frac{\sqrt{6}\sigma}{2\pi b M \mu \sqrt{m} K_M} \propto \frac{1}{M\sqrt{m}\,\mu}, \tag{8}$$

which corresponds to the SNL with total detected photon number $N_{tot} \sim m\,\mu$. Thus, the CBW architecture provides a deterministic $M$-fold enhancement in phase sensitivity with respect to conventional Fizeau-type wavemeters, while remaining consistent with the SNL.

It should be emphasized that the enhancement factor $M$ in Eq. (8) represents a deterministic system parameter – the number of coherently coupled MZIs – and therefore corresponds to a different degree of freedom from the stochastic measurement outcomes that define the Fisher information. In other words, $M$ is not associated with the random variables of the measurement statistics but with the fixed interferometric architecture of the CBW system. By contrast, in N00N-state-based quantum sensing, the parameter $N$ denotes the number of entangled photons participating in the measurement processes and thus directly determines the measurement statistics. Despite this distinction, the CBW-based sensing scheme offers a clear signal-to-noise (SNR) advantage characterized by an effective enhancement factor $MK_M$. When the interferometric order $M$ becomes comparable to the fringe count $K_M$, the resulting scaling approaches $M^2$, arising from deterministic phase multiplication and increased fringe density rather than from quantum entanglement. This architectural amplification also leads to a practical reduction in the required transverse size of the Fizeau wedge optics used to generate the spatial images along the $\zeta$ direction. For a given sensitivity, the transverse dimension of the wedge optics can therefore be reduced compared with that in conventional Fizeau-type interferometers. Specifically, the effective wedge length scales as $L_M = L/M$, indicating that the required transverse extent of the wedge optics decreases inversely with the number of coherently coupled MZIs.

The reduction of the effective wedge length $L_M$ also relaxes the required transverse beam aperture inside the MZI chain. To preserve high fringe visibility, the beam width should remain smaller than $L_M$, preventing phase averaging across the beam profile despite the CBW-induced phase multiplication. In conventional interferometers, increasing spatial fringe density can lead to phase averaging across a finite beam width, which reduces interference contrast. Thus, the CBW-based quantum wavemeter simultaneously provides (i) deterministic M-fold phase amplification to superresolution, (ii) effective sensitivity enhancement approaching $M^2$ when combined with the spatial fringe count, and (iii) a reduction of the required transverse wedge length by a factor of $M$. These combined features enable a compact and high-visibility interferometric wavemeter without requiring quantum-entangled light sources.

Regarding the phase control $\varphi$ in Fig. 1, the phase uncertainty scales inversely with the SNR of the detected interferogram, i.e., $\sigma_\varphi \sim SNR_{amp}^{-1}$, where $SNR_{amp}$ corresponds to the amplitude SNR of $y_k$ in Eq. (5). This phase uncertainty determines the depth resolution used to retrieve the axial information of a target sample, for which the sensitivity is proportional to the SNR. In a typical Fizeau interferometer, improving the $\varphi$-dependent axial resolution therefore requires increasing the detected intensity $\mu$, leading to $\sigma_\varphi \sim \mu^{-1/2}$. The same principle applies to the CBW configuration shown in Fig. 1. Here, the phase $\varphi$ does not need to be continuously scanned; instead, as in conventional Fizeau interferometry, it can be varied discretely with a small number of phase steps. Interpolation between these measurements determines the local gradient of the $\varphi$-dependent interferogram $y_k$ at each lateral position of $\zeta$. The sensitivity of this gradient measurement scales as $\sqrt{\mu}$, reflecting the SNL behavior of the phase fringes. This phase-dependent sensitivity further improves the effective resolution of the CBW interferometer described by Eq. (8). Combined with the spatial fringe multiplication characterized by $K_M$, the overall SNR enhancement can scale as $MK_M$, leading to $M^2$. Consequently, although the CBW architecture can exhibit an effective $M^2$ enhancement in phase sensitivity, this improvement arises from deterministic phase multiplication and increased fringe density, and therefore does not constitute a violation of SNL.

The lower inset of Fig. 1 illustrates a special case with $M=2$, for which $\text{std}\left(\frac{\hat{\Lambda}_M}{\Lambda_M}\right) = \text{std}\left(\frac{\hat{\Lambda}}{\Lambda}\right)/M$. In this configuration, the dummy MZI (red) in Fig. 1 is replaced by a Sagnac interferometer, which provides the practical advantage of reduced environmental noise. The upper inset of Fig. 1 presents a block diagram of the CBW-based quantum wavemeter, where the reference field $E_{ref}$ can be selectively or simultaneously combined with the signal input $E_{in}$, depending on the fringe-counting scheme employed. The anti-symmetrically coupled MZIs can be implemented using bulk free-space optics, fiber interferometers, or integrated optical waveguide platforms such as silicon photonics. If the two output fields are multiplied, the effective resolution becomes twice that given in Eq. (4), without loss of generality. For comparison, conventional high-end wavemeters typically achieve measurement accuracies on the order of parts per million, corresponding to frequency uncertainties of MHz or wavelength uncertainties of approximately $10^{-5}$ nm [32]. Consequently, the CBW effect can contribute to the deterministic sensitivity enhancement discussed above, consistent with the $M$-dependent phase amplification inherent to the CBW mechanism.

Figure 2 presents a proof-of-principle demonstration of the CBW-based quantum wavemeter (see the lower inset of Fig. 1). In the experiment, the phase $\varphi$ is controlled by a piezo-electric transducer driven by a repeated sawtooth voltage ramp. To highlight the CBW mechanism, the wedge optics is omitted for simplicity. The input light is provided by a commercial He-Ne laser operating at $\lambda = 633$ nm, with a linewidth of 5 MHz and an output power of 0.1 mW. For reference, a conventional MZI signal is obtained by tapping one path of the Sagnac interferometer with a thin glass plate, producing the output field $E_{B1}$. The $M = 2$ CBW signal is measured at $E_{B2}$. For direct comparison, the two signals are displayed side by side. As shown, the CBW interferogram ($I_{B2}$; $N = 2$) exhibits twice the fringe density of the reference signal ($I_{B1}$; $N = 1$), as expected (see the moving fringes in the Movie).

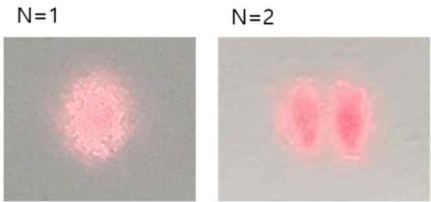

**Fig. 2.** Experimental demonstration of CBW. (a) PZT scan fixed.

Movie. Fringe counts of CBW quantum wavemeter.MOV

In summary, we proposed and analyzed a CBW-based quantum wavemetry scheme that achieves superresolution and enhanced phase sensitivity using an anti-symmetrically coupled MZI architecture. The enhancement factor $M$ in both resolution and sensitivity arises from the $M$th power of the MZI unitary operator acting on the coherent input field, whose photons follow independent and identically distributed statistics. Analogous to N00N-state-based quantum sensing, where $N$ denotes the number of entangled photons, the parameter $M$ in the CBW-based quantum wavemetry represents the number of coherently coupled MZIs. Compared with N00N-state-based supersensitivity, which can surpass the SNL and approach the Heisenberg limit, the CBW-based quantum wavemeter implemented in a conventional Fizeau-type interferometer provides an effective $M^2$ enhancement in phase sensitivity, while the overall noise scaling remains consistent with the SNL. In the CBW architecture, this enhancement originates from deterministic phase multiplication associated with both the longitudinal phase parameter $\varphi$ and the transverse spatial parameter $\zeta$. As a proof-of-principle, we experimentally demonstrated a Sagnac-integrated CBW configuration for $M = 2$, achieving a twofold improvement in fringe resolution compared with a conventional interferometer. The round-trip propagation design used in the CBW quantum wavemeter also provides intrinsic robustness against environmental disturbances, including air fluctuations, temperature drifts, and mechanical vibrations.

**Acknowledgment**

The author gratefully acknowledges Prof. Ben Lee for hosting the author during the sabbatical leave and for kindly providing office space in the department.

**Data availability**
All data generated or analyzed during this study are included in the published article.

**Author contribution**

B.S.H. conceived the idea and solely wrote the paper.

**Funding**

The author acknowledges that this work was supported by the IITP-ITRC grant (IITP 2026-RS-2021-II211810) funded by the Korean government (Ministry of Science and ICT).

**Competing interests**

The author is the founder of Qu-Lidar.